\documentclass[a4paper]{article}
\usepackage{a4wide,url,times}
\usepackage{graphicx}
 \usepackage{times,mathptm}
\usepackage{authors}
\usepackage{rdtjlists}

\usepackage{fancyheadings}

\begin{document}
\title{A functionality taxonomy for document search engines}
\author{\Rik \Erik}
\date{}
\maketitle

% -- "waar naar toe" informatie
% nodig is ook: \usepackage{fancyheadings}
\thispagestyle{fancy}
\setlength{\headrulewidth}{0pt}
\cfoot{\textsl{\footnotesize
  Submitted to: DocEng 2001, ACM Symposium on Document Engineering,
  Atlanta, U.S.A., November 9--10, 2001.}}
% -- Einde "waar naar toe" informatie

% \input{sde01-00}
{\sc Published as:}
\begin{quote}
  R.D.T. {Janssen} and H.A.~(Erik) {Proper}. {A functionality taxonomy for document search engines}. Technical report, Ordina Institute, Gouda, The Netherlands, EU, June 2001.
\end{quote}

\begin{abstract}
  In this paper a functionality taxonomy for document 
  search engines is proposed. 
  It can be used to assess the features
  of a search engine, to position search engines relative to each other,
  or to select which search engine `fits' a certain situation.
  One is able to identify areas for improvement.
  During development, we were guided by the viewpoint of the user. 
  We use the word `search
  engine' in the broadest sense possible, including library and web
  based (meta) search engines.
  
  The taxonomy distinguishes seven functionality areas: an
  indexing service, user profiling, query composition, query execution,
  result presentation, result refinement, and history keeping. 
  Each of these relates and provides services to other
  functionality areas. It can be extended whenever necessary.
  
  To illustrate the validity of our taxonomy, it 
  has been used for comparing various document search engines existing today
  (ACM Digital Library, PiCarta, Copernic, AltaVista, Google,
  and GuideBeam). 
  It appears that the functionality aspects covered by
  our taxonomy can be used for describing these search engines.
\end{abstract}


\section{Introduction}
Searching almost forms an integral part of our life. This paper focuses
on automated searching. Due to the omnipresence of computers,
search engines can be found in almost any application area,
as stand-alone or integrated in other packages, and
in a variety of settings. They can be used
for very different tasks,
ranging from simple fact finding to more complex
decision making and research tasks.
%Search engines are both found as stand-alone engines and integrated
%in other packages.

Examples of stand-alone document search engines are found on 
the web. Their number and capabilities
have increased considerably over the past few years,
mainly caused by the expansion of the 
Internet. There, search is considered to be one of the most visible and
important activities~\cite{brewer01}.
However, the web is not the only place where stand-alone
document search engines can be found. Other examples are
Digital Libraries and databases from large libraries. 
One may even consider a (human) librarian to
be a special type of `search engine'. 

Document search engines integrated in other packages can be found in
e.g.\ Document Management Systems and in Workflow Management
Systems. There, one may imagine a search engine using role or task
bound profiles as a tool for helping workers to find the right
information.
This variety of search engines leads us to define document
search engines in this paper in the broadest sense possible.

The structure of this paper is as follows. 
In the remainder of this section we discuss how we started with this
research and what we consider to be a document.
Section~\ref{sec:user} provides a theoretical perspective on search engines,
where the focus is on the functionality as it may be observed by a searcher.
This is followed, in section~\ref{sec:taxonomy}, by the current version
of our functionality taxonomy.
Before the conclusion, in section~\ref{sec:engines} we
use this taxonomy to position some existing search engines. 
We have taken six search engines
to illustrate how they compare to each other. 

%\subsection{Ongoing research}
\subsection{How we started}
Given the currently existing variety of search engines and their 
ways of deployment, a natural question to ask is `what functionality may
be offered by a search engine?'.
This was the starting point for the development of the functionality
taxonomy as presented in this paper.

Our functionality taxonomy focuses on functionality rather than on the order
in which the functionality is actually provided by a search engine.
Thus, the focus is on \emph{function} rather than on \emph{process},
which acknowledges the fact that two search engines may
cover the same kind of functionality, but use a different process flow 
(strategy) to help users satisfying their information need.
Also, it may be very well possible that the same search engine (as a 
service to the user) offers different 
ways for combining functionality.

The taxonomy as presented here, is a first iteration of ongoing
research, where we both considered theoretical models for functionality
of search engines as well as operational search engines.
Examples of its use are:
\begin{DOTitemnoin}
\item
  to gather, develop or select guidelines on what functionality
  is useful (and to what extent) for applying a 
  search engine in a given situation;
\item
  to determine the (practical and theoretical)
  possibilities of search engines; 
\item
  to determine if a given search engine `fits' some intended purpose;
\item 
  to position different existing (components of) search engines 
  with regard to their functionalities and strategies;
\item 
  to develop standards, based on clearly demarcated areas of functionality,
  for the development of open and standardized search infrastructures.
\end{DOTitemnoin}
On some of these points, the taxonomy does need 
further refinements. Also, quantifications are
required for some of the areas of functionality
to measure the extent and quality 
to which a given search engine provides that functionality.

% Zou het niet leuk zijn om een KPI applicatie te maken op basis van de taxonomy
% om search engines onderling te kunnen vergelijken?

\subsection{Documents --- that what we seek}
The actual information that is sought by searchers who turn to a search
engine is likely to be stored in some \emph{document collection}.
We (just as the organizers of DocEng 2001~\cite{www:sde2001}) 
hold to an expansive notion of documents:
a \emph{document} is a representation of information that is designed to be
read or played back by a person. It may be presented on paper, on a
screen, or played through a speaker and its underlying representation
may be in any form and include data from any medium. A document may be
stored in final presentation form or it may be generated on-the-fly,
undergoing substantial transformations in this process. It can
include extensive hyperlinks and form part of a large web of information.
Furthermore, apparently independent documents may be composed, so that a
web of information may itself be considered a document.

Examples of such documents include: photographs, maps, radio fragments,
movies, a snapshot of the web, and any combinations thereof, such as 
documents containing images and video fragments in any format.

\section{A searcher's perspective on search engines}
\label{sec:user}
%\subsection{The search process}
We have focused our efforts on 
functionality that may be directly observed by users,
independent of the search actor (role). 
This is done on purpose, since a functionality in
the taxonomy may be performed by more than one actor 
where each has its own strengths: for example,
searching can be done by the user 
(the expert considering the information
need, knowing the best what needs to be retrieved), by a librarian (who
is the expert in information retrieval, having access to
resources others have no access to, or are not aware of)
or by an automated search engine
(which may be more easily accessible than a librarian).

\begin{figure}
\begin{center}
  \includegraphics[width=\textwidth]{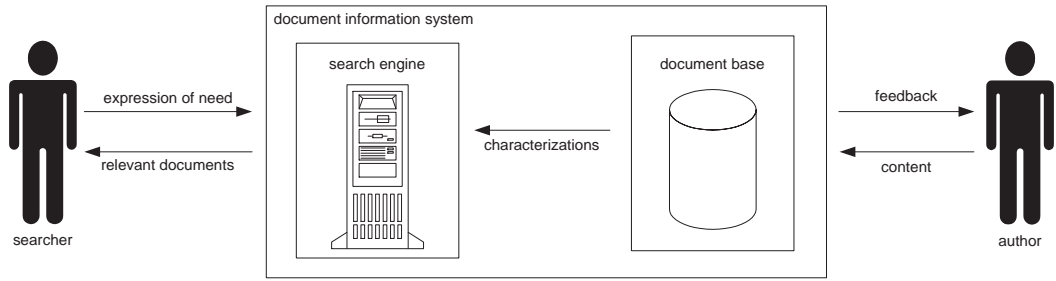}
\end{center}
\caption{The information disclosure paradigm.}
\label{fig:infodisclosure}
\end{figure}

In figure~\ref{fig:infodisclosure}, a conceptual 
architecture of a document information system and its context is given.
An author delivers content to a document base, possibly characterized
using some method. If desired, feedback can be obtained. On the other
side, a searcher expresses an information need to a search engine, which
returns the relevant documents.  We call this the \emph{information
disclosure paradigm}. It is inspired by 
the information discovery paradigm~\cite{proper99} with the addition of
the author's perspective.

For clarity, we make a clear distinction
between search engine functionality, supporting the disclosure of information,
and functionality supporting storage of information.
In this paper, we focus on the left half of 
figure~\ref{fig:infodisclosure}: the searcher and the search engine.  It
is quite likely that a `commercial-off-the-shelf' document information
system offers both document storage and search engine functionality in
one single software package.

From a searcher's perspective, the information disclosure paradigm
highlights some of the key challenges confronting search engines.
Whenever humans have an information need, they try to formulate a
query with the aim of obtaining an answer that best satisfies their need. 
This may sound trivial, but there are several caveats.
To mention a few: formulating a query is not so straightforward, 
some places may be more appropriate than others to go for the information 
(but which ones?), and finding the `right' information between all sources 
returned in an automated search procedure is not easy.
Thus, the initial query will usually only be a crude description 
of the actual documents needed to fulfill the information need.
Therefore, further refinements are done as the search proceeds.
This process is usually called \emph{relevance feedback}~\cite{rijsbergen79}.

The need for information can be caused by a number of 
reasons.
Usually, this is due to some perceived `gap' in the searcher's knowledge.
A gap that may range from being fairly specific such as an answer
to the question `when did Mahatma Gandhi die?', to very broad, 
such as learning about `the world of Mahatma Gandhi'.
A specific need can usually be satisfied by a small collection of facts,
while a broad need usually requires a wider variety of facts.
During the search process users may learn more and more
about their knowledge gap, and may thus discover aspects of
this gap they were initially not aware of.
This means that the actual information need of a user may evolve
as they are exposed to new information.

%Essentially, from a searcher's point of view, the key functionality
%that is needed from a search engine, is finding the right
%documents (or fragments thereof) that fill the user's knowledge gap.
%There are three issues to play a central role:
%formulation of queries, characterization of documents, and
%selection of relevant documents (or fragments). 
%
%The formulation of queries is not a trivial task.
%involves two important
%issues.
%First of all, it requires some formal language in which to express
%the query.
%Secondly, a precise formulation of the \emph{true} information
%need is required.
%Although it is acknowledged in e.g.\ the Cranfield tests
%\cite{cleverdon91} that users have
%difficulty in expressing their information need in a formal language,
%the fact that searching for information is more of an interactive
%process of learning, clarification and discovery is quite often not taken 
%into account.

Traditionally, the quality of a search engine is measured in terms of
precision and recall \cite{rijsbergen79}.
These only measure the ability of a search engine
to effectively execute a query. This seems less suitable for assessing
interactive systems, because in many interactive
settings, users require only a few relevant documents and do not care
about high recall~\cite{lagergren98,bruza00}.
Useful other metrics to define successful queries
include: time required to learn how to use the
system, time required to achieve goals on benchmark tasks, or the error rate.

\section{Functionality taxonomy}
\label{sec:taxonomy}
\begin{figure}
\begin{center}
   \includegraphics[width=\textwidth]{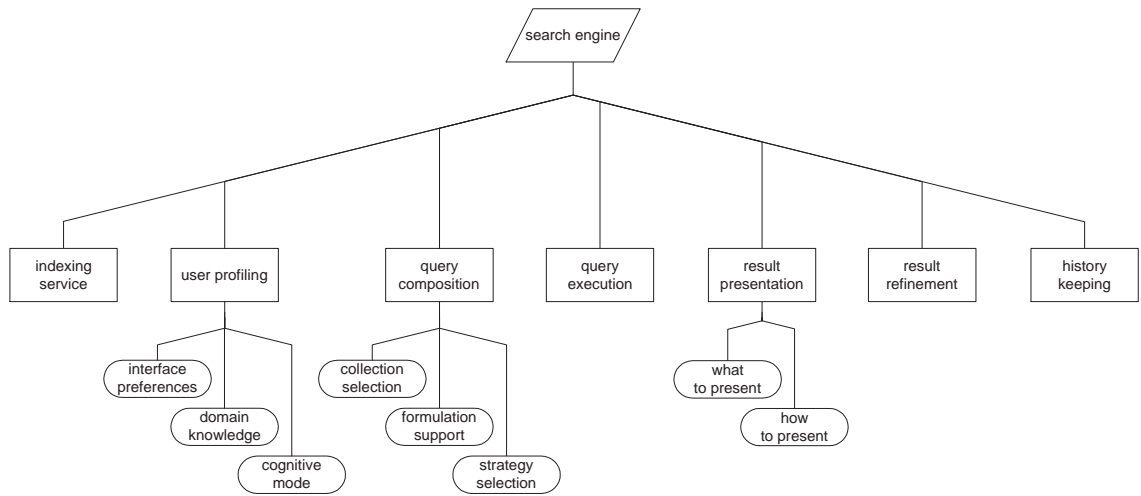}
\end{center}
\caption{The functionality taxonomy.
A tree-like representation has been used to emphasize
functionality rather than a particular order.}
\label{fig:queryflow}
\end{figure}

Functionality of search engines can be distinguished at different levels of 
detail.
In the initial version of our taxonomy, we distinguish seven general
functionality areas, represented by rectangles in 
figure~\ref{fig:queryflow}.
These seven functionality areas
to fulfill information needs are: an
indexing service, user profiling, query composition, query execution,
result presentation, result refinement, and history keeping. 
Each of these functionality areas relates and provides services
to other functionality areas.
In subsequent versions of the taxonomy, more detailed levels of
functionality and quantifications thereof will be added.

The general functionality areas
user profiling, query composition, and result presentation,
have been further subdivided into different aspects. 
These aspects are intended to fully cover the functionality area
directly above it. 
For example, for the user profiling functionality area, 
we consider the three aspects interface preferences, domain knowledge,
and cognitive mode. 
In subsequent versions of the taxonomy, additional aspects are likely
to be identified.

The remaining functionality areas have not been subdivided yet; not because 
these are not considered to be important, but rather since we have decided
to focus initially on functionality directly observed by a searcher.
These areas are briefly discussed in section~\ref{sec:otherfuncts}. 

%We can make a distinction between two classes of queries: (1)
%  the `you know what you want' query, where the challenge is how to formulate
%  the query to obtain the answer to the query; and (2) the
%  `you do not know what you want' query, which is more like an explorative
%  query on an `abstract' level, where the user more likely wants to be
%  guided in some way to find the answer to the query.

\subsection{User profiling}
User-interface design for information retrieval tasks is a non trivial
task with its own set of challenges, such as diversity of the user
communities broadly accessible resources like the web or library systems
propose to serve~\cite{shneiderman97}. The successfulness of the search
is influenced by the variety in (cultural and
professional) background, experience, and knowledge.
It is therefore that for search engines user profiling
is an important functionality.

Currently, we distinguish three aspects which together comprise a user profile:
\begin{DOTitemnoin}
\item
  interface preferences,
\item
  domain knowledge, and
\item
  cognitive mode.
\end{DOTitemnoin}
These aspects are explained in more detail below.
The user profile is concerned with the way in which a searcher wants to use
a search engine,
both for the long term (some user preferences) and for the short term (some
cognitive modes), both dynamically (domain knowledge which changes due
to queries and their results) and more-or-less statically (some user
preferences).
Ideally, user profiles should be portable between search engines, so
that searchers can use different search engines more easily.

\subsubsection{Interface preferences}
% \paragraph{What user interface?}
Interface preferences are concerned with the style of user interfacing
that is to be used. 
This can depend on many things, such 
as the preferences or the information need of the searcher. 
Possibilities include command language, form fill-in, selection from
a menu, and
natural language.

% \paragraph{How to interface?}
Depending on the user interface, there are several ways to 
interact with the searcher. For instance, the searcher 
might prefer the keyboard, or rather use a pen-pad to sketch a shape he
wants the search engine to retrieve. Other possibilities are
speech input and output, or files to interface with a statistical
software package.

% \paragraph{User experience with search engine.}
Searchers with differing levels of experience may be supported by appropriate
variations in the user interface. There are two variants
of experience between searchers:
differences in experience of the searcher with the search engine (both
frequency of use and the skill of the searcher with a particular search
engine) and
differences in domain knowledge of the searcher (see the next section).
It can be argued~\cite{hoelscher99} that both influence the effectiveness 
of the search. 
An example of adapting the user interface to the skill of the searcher 
with a particular search engine is given in~\cite{shneiderman92}:
\begin{DOTitemnoin}
\item
    first-time searchers need an overview to understand the range of
    services plus buttons to select actions;  
\item
    intermittent searchers need an orderly structure, familiar landmarks,
    reversibility, and safety during exploration;  
\item
    frequent searchers demand shortcuts or macros to speed repeated tasks
    and extensive services to satisfy their varied needs.
\end{DOTitemnoin}

\subsubsection{Domain knowledge}
% \paragraph{User domain knowledge.}
Searchers with different domain knowledge need different answers to their
queries. For instance, a searcher wanting to learn a new computer language
who is a novice in the area of computer
programming, needs an introductory text and a gentle introduction to the
computer language itself, while someone who is already able to program in 
several different computer languages may only need a reference book.

Adapting search engines to searchers with varying degrees of domain knowledge 
is a more difficult task than adapting search engines to the experience 
of the searcher with that particular search engine, since in the former case
there has to be a way for the system to `learn' 
the level of expertise from the searcher. 
This is dynamic knowledge, since
the domain knowledge of a searcher
is likely to change while reading the answers returned from a query.

\subsubsection{Cognitive mode}
% \paragraph{User information need.}
Searchers may be in various cognitive modes when they turn to a search
engine~\cite{proper99,proper01}. 
One may think of examples such as able to learn or not, happy or not,
in a hurry or not, tired or not, or
willing to try something new or rather use somewhat more
familiar. 
Not many (if any) search engines existing today are sensitive to
a searcher's cognitive mode.

Automatically detecting a searcher's cognitive mode is not an easy
thing to do. 
Using contextual information such as the time of the day or the
task at hand, may help in resolving this mode~\cite{ram94}.
Knowing it may allow a search engine to better tune its 
activities to the searcher, and subsequently, to improve the satisfaction of
the searcher. Besides, it also determines the way results
should be presented to the searcher. 
Another example aspect of a cognitive mode is the overall search
goal, i.e.\ is the goal to obtain an answer:
\begin{DOTitemnoin}
\item
  to a precise question (one answer exists and can be given) such as `When
  did Mahatma Gandhi die?'. % We call this: \emph{exact query}.
\item
  to fill a gap in the searchers knowledge (many answers are possible, maybe
  there is no definite answer),
  such as `What were the circumstances in which 
  Mahatma Gandhi died?' % We call this: \emph{information discovery}.
\item
  to learn about a certain subject on the basis of a
  `recommended concise reading list',
  such as `Tell me about the world
  of Mahatma Gandhi'. % We call this: \emph{information coverage}.
\end{DOTitemnoin}
Each of these search goals is more complex than its predecessor and 
requires more functionality from the search engine.
%For example, with information coverage
%one would want the system to extract the exact knowledge fragments from
%the relevant documents, so that users would not have to read 
%entire documents.
%
%This requires the system to have some form of natural language
%parsing and understanding. Additionally, the system should have a method
%for determining what the user already knows, thus
%preventing to present material already known.

\subsection{Query composition}
The query composition functionality assists searchers in formulating their
information needs in terms of a query.
We distinguish three aspects which form the query composition 
functionality (to be discussed in the following sections):
\begin{DOTitemnoin}
\item
  collection selection,
\item
  formulation support, and
\item
  strategy selection.
\end{DOTitemnoin}

It can be observed that existing search engines have a wide
diversity in user-interface, query language, etc., 
and that no `standard' exists (yet) to
prevent inconsistencies between different search services. 
Also, often the processing of the search engine on the input of the searcher
and how the search engine executes the query is unclear.

This is important, since some searchers may use more than one 
search engine, and by unclarity what exactly 
happens after they have entered their query,
lower performance, uncertainty, mistaken assumptions, and failures to
find relevant documents, may occur.
This is illustrated by the following example 
(from~\cite{shneiderman97}), where the search string `Hall effect'
could produce (among many other possibilities) a:
\begin{DOTitemnoin}
\item
    search on the exact string `Hall effect';
\item
    case-insensitive search on the string `hall effect';
\item
    probabilistic search for `Hall' and `effect';
\item
    probabilistic search for `Hall' and `effect', with higher 
    weights if `Hall' and `effect' are in close proximity;
\item
    error message indicating missing \textsc{and/or} 
    or other operators/delimiters;
\item
    Boolean search on `Hall' \textsc{and} `effect';
\item
    Boolean search on `Hall' \textsc{or} `effect'.
\end{DOTitemnoin}
In the following sections, the aspects from the query composition
functionality are discussed.

\subsubsection{Collection selection}
% \paragraph{Which sources?}
Collection selection allows the searcher to
choose which collections to use with the current search engine. 
This makes the user aware
of the different collections available. 
An option is to offer some predefined set.

On the web, queries can be sent to many search engines. Libraries
may have several different collections (or databases)
with each its own search engine. For example, Pica~\cite{pica}, a
company in the Netherlands, offers database services
such as NCC (the Dutch central library catalogue) and OLC (On Line
Contents)\footnote{%
  NCC contains bibliographic references and the locations of approximately 14
  million books and almost 500,000 periodicals in more than 400 libraries
  in the Netherlands. 
  OLC contains references to all articles that appear in
  almost 15,000 current periodicals in all fields of science.
}.
Another service, called PiCarta (see section~\ref{sec:picarta})
to search these and others databases at the same time, is offered additionally.

\subsubsection{Formulation support}
% \paragraph{How to start the query?}
Formulation support helps searchers to formulate their query.
One of the aspects is that searchers need a clue to decide how to
start the query process. Studies show that searchers tend to start out
with very short queries, inspect the results, and then modify those
queries in an incremental feedback cycle~\cite{anick94}.
According to~\cite{baeza99}, four main types of starting points for
queries can be distinguished.
\begin{DOTitemnoin}
\item
  Lists: a query is started with a long list of collection names and the
  searcher is required to guess which one is of interest. From these kinds
  of lists, frequent searchers may make their own list of `favorites' or
  `bookmarks'.
\item
  Overviews: a searcher can use an overview to select or eliminate
  the topic domains represented in the various collections. Such an
  overview can be used to get started, directing searchers to general
  neighborhoods, after which they can navigate using more detailed
  descriptions. This starting point is different from lists in that it
  may offer grouped overviews.
  One may consider various types of overviews:
  \emph{category hierarchies}, often associated with a certain
  discipline such as \textsc{medline} or \textsc{acm}, automatically
  derived by unsupervised \emph{clustering techniques} on the text of the
  documents attempting to extract overall characterizing themes, 
  and derived using \emph{co-citation analysis} on connections or links 
  between the different documents of a collection. Other possibilities
  are graphical depictions of bookshelves or piles of books.
\item
  Examples: the searcher is shown a general query template which can be
  modified to construct a description of what they want. 
  Next, the system shows an example of the kind of information available
  that matches the description. This may be called \emph{retrieval by
  reformulation}. Other possibilities are \emph{wizards}, which provide
  the searcher a step-by-step shortcut through the (usual) sequences of 
  menu choices possible, and a \emph{guided tour}, leading a searcher
  through a sequence of navigational choices through hypertext links,
  presenting the nodes in a logical order.
\item
  Automated collection selection: this requires both eliciting the
  information need from searchers and understanding which needs can be
  satisfied by which collections. Another possibility is to create a
  representation of the contents of information collections and match this
  representation against the query specification, or to send the query
  to multiple collections and combining the results from the various
  systems (often used by web-based meta search engines).
%  All of these are areas wide open to research.
\end{DOTitemnoin}
%
% \paragraph{Which fields?}
Queries can relate to all fields of the database, or to just a few ones. On
basis of experience, the searcher may choose to query on certain fields. For
instance, when a searcher wants to query for journal articles written by a certain
author, he may choose to only access the author-field 
and not the editor-field to prevent erroneous matches.

% \paragraph{What to search?}
% The searcher uses the searcher interface and the method to interface
% as described in the previous paragraphs, to specify what values to search. The
% fields and sources have been specified in previous steps.

\subsubsection{Strategy selection}
% \paragraph{How to search?}
The search engine may apply several strategies when executing
a query. In the following paragraphs
we discuss some strategies we have encountered while investigating literature
and search engines existing today. 

\paragraph{Matching options.}
There are several different ways to match a query to the
characterization of a document. 
Two of the best known are best-match and exact-match (Boolean) search. 
In addition, several variations of fuzzy-matching and 
partial-matching\footnote{For example, with partial-matching
  searching for `biology' also retrieves `sociobiology' 
  and `astrobiology'.} 
exist.

The study in~\cite{byrd00} has shown that
most common searchers seem to have a preference for best-match
systems, while expert searchers seem to prefer exact-match
by an overwhelming margin.
This seemed to be the case since
it is easier to explain searchers why a Boolean system did or did not
retrieve a given document, regardless of its actual relevance. 
This cannot be said for a
best-match system. For example, with the query `Monarch \textsc{and} 
butterfly', a Boolean system will retrieve all documents that use both
words and no other documents, and it will therefore be obvious why any given
document was or was not retrieved.
When this is compared
to the query `Monarch butterfly' for a best-match
system, it may be that not only the system likely is 
to retrieve documents that discuss
either queens and kings, or butterflies in general, without mentioning
Monarch butterflies; but ---worse--- it may even be possible that 
some of these documents rank above any document that really does
refer to Monarch butterflies.

\paragraph{Query transformations.}
There are many kinds of transformations a search engine may apply on a query
after the searcher has entered the query.
Examples of common transformations are: 
\begin{DOTitemnoin}
\item
  stemming of words, that is dealing with the conversion of
  words to their presumed roots, e.g.\ `blacker', `blackest', 
  and `blacks' may all be converted to `black';
\item
  case insensitivity;
\item
  removal of stop words, whereby the system automatically
  ignores words that are
  assumed to be so common as to carry little information useful for
  distinguishing relevant documents from non-relevant ones;
\item
  soundex expressions, where the system also queries for words which sound
  like the one specified in the query;
\item
  associating weights to query words depending on the position 
  in the query; 
\item 
  coreference resolution, where the system uses variations in a phrase
  to refer to an object~\cite{ogilvie00}.
  One type is the use of pronouns to refer to a named-entity, 
  for instance, if `Mahatma Gandhi' is mentioned in some query, the 
  object may later
  be referred to as `he' in the same query. 
\end{DOTitemnoin}
These kinds of transformations may confuse searchers if systems do not
give indication which are being applied. Note that these kinds of
transformations may also be applied when deriving
characterizations (such as keywords) from documents. 

Coreference transformation is a challenging
task~\cite{grishman97} as it requires some degree of natural language parsing.
It can be done on one document in isolation or across documents.
There are several types of coreferences, such as 
the use of pronouns mentioned before.
Another type consists of variations
on named-entities or noun runs. That means that 
`Mahatma Gandhi' is referred to as `Mr. Gandhi', or that
phrases such as `the pacifist' or `the person committed to poor people'
are used. Acronyms are another type. 

\paragraph{Multiplicity of queries.}
Querying a system only one time is different from asking the same query at
different time intervals. In the latter case, the searcher probably wants to
be kept informed about developments in a certain area or wants to be
informed whenever new documents on the same subject are being added to
the information collection.

\paragraph{Time of execution.}
A searcher may want to submit a query to be executed at a later time for
various reasons, such as that the time expected for completion may 
be too long, or that queries at night may be less expensive (i.e.\ cost
less money) to execute.

\subsection{Result presentation}
The result presentation functionality determines the way the
results are displayed. There are two aspects:
\begin{DOTitemnoin}
\item
  what to present (e.g.\ which parts and which fields), and
\item
  how to present. 
\end{DOTitemnoin}
%
% \paragraph{How to display?}
For the latter, results can be
presented using visualization techniques or they can be clustered. 
The use of clustering is
based on the cluster hypothesis of information retrieval:
`closely associated documents tend to be relevant to the
same requests'~\cite{rijsbergen79,leuski00a}.
It appears that searchers are most often confused 
by not being given any clue in which order the results are
presented. Options are, for example, to specify
layout and sequencing (alphabetically, chronologically,
relevance ranked, etc.). 
%
% \paragraph{Which visualization techniques?}
Instead of presenting the searcher an ordered list with results, results
can be presented visually (an area wide open to research).
Numerous studies suggest that techniques
such as topic-based grouping of similar documents
(`how do the resulting documents relate to each other') 
are a better way of organizing the retrieval results. 
An overview of related work
on clustering and document visualization can be found in 
e.g.~\cite{baeza99,leuski00b}.

\subsection{Remaining functionality areas}
\label{sec:otherfuncts}
This section briefly discusses the remaining functionality areas as 
mentioned in figure~\ref{fig:queryflow}. In future reports these will 
be subdivided in
several aspects and discussed in more detail.
In the section~\ref{sec:engines} the taxonomy from this section
will be used to assess selected search engines.

\subsubsection{Indexing services}
Indexing services offer indexes on the
documents in the document collections. These indexes can be generated
on the fly or in advance, automatically or by hand. Search engines may
generate these themselves or obtain (and possibly combine) 
them from the various collections and information sources
they have access to. Indexing services may also provide access to thesauri.

To generate indexes automatically numerous document recognition and
interpretation techniques may be used. 
As these techniques are outside the scope of this paper,
they are not discussed. 

\subsubsection{Query execution}
This functionality is concerned with the actual execution of the query.
The basic result set may be post processed to become more useful to the
searchers, for example to remove duplicate answers obtained from separate search
engines or to remove all but the top-10 answers.

% \paragraph{How to activate the search?}
Queries may be activated explicitly or implicitly. 
Typical is to have searchers click on a `search' button to
initiate the search and then wait for the results. Another
alternative is that of dynamic queries:
the result set is continuously displayed and updated during the query
process. This approach most often requires 
high bandwidth and (for large databases) very rapid processing. Some
advantages are that searchers can broaden, narrow, or
refocus their query several times in as many seconds.

% \paragraph{Inspection of internal query representation.}
The search engine interprets the query and may respond with some abstract form 
of the query, allowing the searcher to modify it. As this might be
especially useful for the expert searcher, this step may be
hidden from the non-expert user.

\subsubsection{Result refinement}
The searcher may choose to
reformulate the query to obtain different or more to
the point answers. This is facilitated by the result refinement
functionality.
Relevance feedback is used to
specify, for example, which query results resembled the
searcher's need and which did not. 
This kind of feedback can be used in a new or modified query to the
search engine.

\subsubsection{History keeping}
% \paragraph{What history to keep?}
History is important when queries are rephrased multiple times and
submitted again to the search engine, or when documents returned from the
query are explored. The history is useful because the searcher is able to
know where he/she was going to and from where he/she came. History keeping
can be done with e.g.\ certain visualization techniques.

\section{Search engine evaluation}
\label{sec:engines}
We evaluated different search engines to assess the `effectiveness' of
our taxonomy as well as to obtain an overview of what types of
functionality was provided. The following search engines have been
evaluated:
ACM Digital Library, PiCarta, Copernic, AltaVista, Google, and GuideBeam.
In table~\ref{tab:comparison} the overall evaluation results are shown.
Some aspects are discussed shortly in the next sections, however, for
the direct comparison with the taxonomy one is referred to the table.

For this paper, due to time and space limitations,
we have chosen to only use a Boolean scoring approach.
A cross in the results table means that the
search engine provides some of that functionality. 
We did not show
to what extent this functionality is provided. An empty cell simply
means that the functionality is not provided. Depending on one's search
goal, that may be `good' or `bad'.
As mentioned in the introduction, one of the next steps should
be to define, for each area of functionality in the taxonomy, suitable
quantifications that allow to quantify to what extent and quality
a search engine offers this functionality.

\begin{table}
\begin{center} \small
  \tabcolsep=0.5\tabcolsep
  \def\X{{\normalsize $\times$}}
  \begin{tabular}{ll|c|c|c|c|c|c}
     \multicolumn{2}{l|}{search engine}          & ACM DL & PiCarta & Copernic & AltaVista & Google & GuideBeam \\
     \hline
     indexing service    &                       & \X     & \X      &          & \X     & \X     &     \\
     \hline
     user profiling      & interface preferences &        &         &          & \X     & \X     & \X  \\
     \cline{2-8}
                         & domain knowledge      &        &         &          &        &              \\
     \cline{2-8}
                         & cognitive mode        &        &         &          &        &              \\
     \hline
     query composition   & collection selection  &        & \X      & \X       &        &              \\
     \cline{2-8}
                         & formulation support   & \X     & \X      & \X       & \X     & \X     & \X  \\
     \cline{2-8}
                         & strategy selection    & \X     & \X      & \X       &        &              \\
     \hline
     query execution     &                       & \X     & \X      &          & \X     & \X     &     \\
     \hline
     result presentation & what to present       & \X     &         & \X       & \X     & \X     &     \\
     \cline{2-8}
                         & how to present        & \X     & \X      & \X       & \X     & \X     &     \\
     \hline
     result refinement   &                       & \X     & \X      & \X       &        &              \\
     \hline
     history keeping     &                       & \X     & \X      & \X       &        &              \\
  \end{tabular}
\end{center}
\caption{The evaluation of search engines according to our taxonomy.}
\label{tab:comparison}
\end{table}

\subsection{ACM Digital Library}
%ACM, the Association for Computing Machinery, is an international
%scientific and educational organization dedicated to advancing the arts,
%sciences, and applications of information technology. 
%They have the ACM Digital Library \cite{www:acmdl}, 
The ACM Digital Library \cite{www:acmdl} offers on line access to
a vast resource of bibliographic information, citations, 
and full-text articles.
The library offers an indexing service by browsing the
ACM journals and magazines and the ACM proceedings by subject, sponsor
or series. Formulation support can be obtained by browsing e.g.\ the ACM
proceedings. One can also specify which terms to use for which fields
(title, full text, abstract, review, and index terms), and there is a
separate facility for querying on author names.

For the strategy selection there are many possibilities.
In formulating the query the usual expressions like 
like \textsc{and, or, not}, and \textsc{near} can be used or more uncommon ones like
fuzzy, synonym, soundex, and stem searches. 
After the execution of the query (where one can inspect the internal
query representation) one has the option to choose what to
present (brief and full listings) and how to present (all search results
or a limited number). 
Results can be ordered by score, publication title or publication date. 

Results can be refined using the same operators available as the first
query. Each query can be saved in what is called a `personal binder'. If 
searchers want, they can be notified with email on updated query results.

\subsection{PiCarta}
\label{sec:picarta}
PiCarta~\cite{pica} is a service offered by Pica, 
and gives (in the basic form) on line access for members to 
NCC (the Dutch central library catalogue), OLC (On Line
Contents), and NetFirst (a catalogue for Internet resources). The searcher 
can choose which of these collections to query and on which types of material
(books, articles, letters, audio visuals, printed music, etc.).

Formulation support is offered for all fields in the
database such as the usual author and title fields, 
but also on several codes and other specialist fields such as
\textsc{isbn} and \textsc{issn} number, 
Library of Congress number, and on accepted medical terms.
Query terms can be combined using the usual Boolean operators and using
wildcard and proximity operators. Stop words are removed.

PiCarta has only one method for result presentation: all fields in
lists of 10 items. This list can be sorted either on relevance or on
year of publication. Refinement can be done by combining the results
from different queries or by doing a reduce, enlarge or except search on
the query results. Query results are saved for one session only.

\subsection{Copernic 2001 Pro}
Copernic 2001 Pro \cite{www:copernic} is 
a meta search engine for web pages,
installed on the computer of the searcher. It offers access to
approximately 1000~search engines and 93~specialized search categories.

Formulation support is provided by using the specialized search
categories. One can use the usual Boolean operators, and there is
support for scheduled search updates. Results can be emailed.
There is
no query execution since the query is sent to the different web search
engines, of which Copernic only combines the results.

Result presentation allows searchers to choose to show all documents
or only the new, downloaded, refined or check marked documents.
Results can be sorted using title, score, web address, or
date found.
Refinement of the results can be done by using Boolean operators.
Another form of refinement consists of downloading the query results
and removing dead links.

\subsection{AltaVista}
AltaVista \cite{www:altavista} is one of the `oldest' search engines for
the web. It uses a keyword based indexing and querying mechanism.

Searchers can specify several preferences, such as the interface language,
filtering of adult content and several display preferences.  AltaVista
allows searchers to focus their searches on a specific website, domain
or region, rather than on the whole web.  To help searchers in finding
their way, AltaVista also offers a topic-based web directory. This
directory is of a similar structure as the one provided by the 
Open Directory Project~\cite{www:dmoz}, and 
seems to be maintained by AltaVista themselves.

\subsection{Google}
Google~\cite{www:google} is based on a technology called
PageRank~\cite{brin98}. This technology 
relies on text-matching techniques as well as on mechanisms to
rank the quality of sites. This mechanism 
is based on the link structure `surrounding' a page: the more it is
referenced, the higher the value (and, probably, the higher the quality).

Searchers can specify several preferences, such as the interface language,
filtering of adult content, preferred language for searched documents and
display preferences. Google allows searchers to focus their
searches on a specific website, rather than on the whole web. To help
searchers in finding their way, Google also offers a topic-based web
directory, based on the Open Directory Project.

\subsection{GuideBeam}
GuideBeam~\cite{www:guidebeam} is a meta-search engine that focuses on
helping searchers to formulate a query. The resulting query is executed using
one of three search engines (one of which is Google).

GuideBeam is basically a research prototype that uses a
query-by-navigation strategy to aid searchers in
formulating their information needs~\cite{bosman98}.
The underlying idea is to present
searchers with an abstract presentation, in terms of noun-phrases, of the
information that is available to them. This allows searchers to clarify
their information need in a process where they first specify some
keywords to the system. These keywords provide the system clues on
the searchers actual information need. Rather than, like most search
engines do, immediately returning large result sets, the
system continues by returning suggestions on possible refinements of
these keywords in terms of more complete noun-phrases. This refinement
(and enlargement) process continues until searchers are satisfied with the
reformulation of their information need. This result is used to
compute the real result set using the selected query engine.

\subsection{Search engine evaluation conclusion}
The search engines in the sections above were chosen because some give
access to large information collections where new documents are only added
whenever they fulfill some standard (ACM Digital Library and PiCarta),
where others give access to large information collections (the web)
where anyone can add new documents (the others).

Copernic and GuideBeam were chosen because both are meta search engines,
the first installed on the computer of the user and the latter directly
accessible from the web. AltaVista is one of the `oldest' search
engines, and Google was chosen because GuideBeam uses it as search engine. 

Another reason for choosing these six search engines with very different
design goals was to illustrate that the
various functionality aspects present in our taxonomy are present in
today's search engines (with the exception of `domain knowledge' and
`cognitive mode', which are candidates for further research).

As can be observed from table~\ref{tab:comparison} AltaVista and Google have
the same score. 
This is caused by the Boolean scoring approach.
However, some searchers are likely to appreciate one above the other, probably
because of the presentation of the results or differences in
perceived quality. 
A scoring mechanism using quantifications could have captured these
differences. In a future paper we will present such a scoring approach.

\section{Conclusion}
In this paper, we have proposed a functionality 
taxonomy for document search engines. 
This taxonomy emphasized functionality from the viewpoint of the searcher.
The word `search
engine' in this paper is meant to be used
in the broadest sense possible, so that it includes web
based (meta) search engines, library search engines, and so on.

The taxonomy distinguishes seven functionality areas:
indexing service, user profiling, query composition, query execution,
result presentation, result refinement, and history keeping. 
It has been set-up in a hierarchical way, which means that more detail 
can be added as needed. 
  
Next, this taxonomy was used for comparing various search engines. We
have evaluated several search engines existing today:
ACM Digital Library, PiCarta, Copernic, AltaVista, Google,
and GuideBeam. 
It appears that the functionality aspects covered by
our taxonomy can indeed be used for describing these search engines.
However, further refinement of the taxonomy is needed, as well as the
quantifications for measuring the extent (and quality) in
which a search engine provides a certain functionality.

The taxonomy in this paper may also be viewed as the
starting point of an architecture for an open and standardized
search infrastructure. 
Interesting research areas are e.g.\ portable user profiles, common
development frameworks for search and retrieval engines, and the
development of (components of) search engines.

%[beetje vage laatste zin] [sowieso een beetje vage samenvatting]
%
%Hebben we nu een antwoord op:
%\begin{itemize}
%\item
%  voor wie is deze taxonomie?
%\item
%  wat kan je ermee?
%\item
%  wat is nieuw tov bestaande systemen?
%\item
%  kader, doel, relevantie?
%\item
%  open en closed information spaces?
%\item
%wat hebben we nu geleerd?
%\item
%hoe was het proces?
%\item
%hebben we ons doel bereikt?
%\end{itemize}

\begin{small}
\bibliographystyle{alpha}
% plaats onderstaand statement in sde01.bbl als de referenties dichter
% bij elkaar moeten (op 0pt afstand). Op deze plek werkt het niet.
% \setlength{\itemsep}{0pt}
\bibliography{all}
\end{small}

\end{document}